\documentclass{article}

\usepackage{graphicx}
\usepackage{enumitem}
\usepackage{amsfonts}
\usepackage{latexsym} 
\usepackage{epsfig}
\usepackage{amssymb}
\usepackage[leqno]{amsmath}
\usepackage{array}
\usepackage{color}
\usepackage{xcolor}
\usepackage{algorithm}
\usepackage[noend]{algpseudocode}

\newcommand{\pf}{\noindent{\em Proof: }}
\newcommand{\epf}{\hfill\hbox{\rule{3pt}{6pt}}\\}

\newtheorem{lemma}{Lemma}

\newtheorem{theorem}{Theorem}

\newcommand{\Z}{{\mathbb Z}}

\newcommand{\cQ}{{\mathcal Q}} 
 
\newcommand{\cS}{{\mathcal S}}
\newcommand{\cT}{{\mathcal T}}

\begin{document}

\title{Arboreal networks and their underlying trees}

\author{
	{\bf Katharina T. Huber} and {\bf Darren Overman}\\
	School of Computing Sciences,\\
	University of East Anglia,\\
	Norwich, NR4 7TJ, UK\\
}
\date{\today}

\newpage

\maketitle

\begin{abstract}
Horizontal gene transfer (HGT) is an important process in bacterial evolution. Current phylogeny-based approaches to capture it cannot however appropriately account for the fact that HGT can occur between bacteria living in different ecological niches. Due to the fact that arboreal networks are a type of multiple-rooted phylogenetic network that can be thought of as a forest of rooted phylogenetic trees along with a set of additional arcs each joining two different trees in the forest, understanding the combinatorial structure of such networks might therefore pave the way to extending current phylogeny-based HGT-inference methods in this direction. A central question in this context is, how can we  construct an arboreal network? Answering this question is strongly informed by finding ways to \textit{encode} an arboreal network, that is, breaking up the network into simpler combinatorial structures that, in a well defined sense uniquely determine the network. In the form of triplets, trinets and quarnets such encodings are known for certain types of single-rooted phylogenetic networks. By studying the underlying tree of an arboreal network, we compliment them here with an answer for arboreal networks.
\end{abstract}

\section{Introduction} \label{intro}

Evidence suggests that bacteria living in distinct ecological niches such as the human gut and the human skin can exchange genetic material via horizontal gene transfer \cite{JACKN19}. Until now, the standard approach in phylogenetics to model this phenomenon has been to assume a common phylogenetic tree for the species of interest called a species tree and to then explain the evolutionary signal  conveyed by a gene tree within the assumed species  tree \cite{BAK12,JACKN19}. To do this, a cost is generally assigned to permissible evolutionary events such as speciation, duplication, and loss and the task is then to find an embedding of the gene tree within the species tree that is optimal in terms of that cost function. 
Although much research has gone into extending and refining this approach  by, for example, assuming additional evolutionary events such as incomplete lineage sorting, not much is known for the case where a species tree cannot be found with high enough confidence or where the information that the exchange of genetic material between niches is important and that this information should therefore be preserved.

\begin{figure}[h]
	\begin{center}
		\includegraphics[scale=0.33]{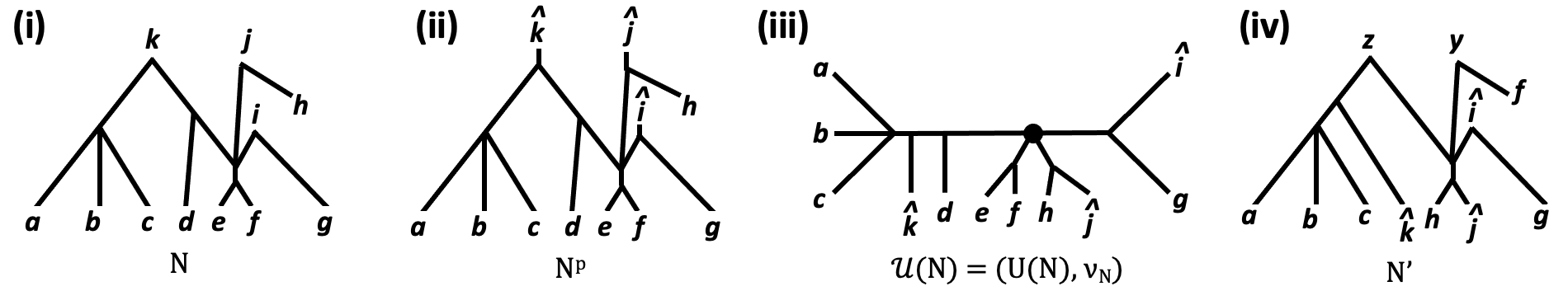}
	\end{center}
	\caption{
		 (i) An arboreal network $N$ with leaf set $X=\{a,\ldots, h\}$ and root set $\{i,j,k\}$. (ii) The 
		 planted version $N^p$ of $N$ with root set $\{\hat{i},\hat{j},\hat{k}\}$.(iii) The augmented tree $\mathcal U(N)=(U(N),\nu_N)$ with leaf set  $X\cup \{\hat{i}, \hat{j},\hat{k}\}$ associated to $N$ where $\nu_N$ is the map that assigns $\bullet$ to the indicated vertex and $\circ$ otherwise. For clarity purposes, only vertices that are assigned the value $\bullet$ under $\nu_N$ are indicated.  (iv)  With $z$ playing the role of $\hat{d}$ and $y$ playing the role of $\hat{e}$, a further arboreal network on $\{a,b,c,\hat{k},h,\hat{j},f,g\}$ that induces $\mathcal U(N)$.  
		\label{fig:example}
	}
\end{figure}

To start filling this gap, multiple-rooted phylogenetic networks such as the one depicted in Figure~\ref{fig:example}(i) 
have been introduced and studied in the literature in the form of Overlaid Species Forests \cite{HMS22a}, forest-based networks \cite{HMS22b,HvIMS25,LM25} and arboreal networks \cite{HMS24}. Although distinct combinatorial objects all of whose set of leaves is a pre-given set $X$ of organisms, they all  can be thought of as a forest $S$ of rooted phylogenetic trees (each representing, for example, the evolutionary past of a set of bacteria inhabiting an ecological niche)  along with a set $A$ of additional arcs (each representing, for example, a horizontal gene transfer event) such 
that each arc in $A$ joins two distinct  trees in $S$.

Arboreal networks $N$ are at the centre of this paper and, in addition to being multiple-rooted,  enjoy the property that their {\em underlying graph}  i.e. the
graph obtained from $N$ by ignoring directions and suppressing all {\em roots} (i.e. vertices with indegree zero and outdegree two or more) that have become vertices of degree  two are unrooted phylogenetic trees in the usual sense (see Sections~\ref{sec:arboreal} and \ref{sec:prelims} for formal definitions, respectively).
Although relatively simple looking structures, arboreal networks enjoy interesting combinatorial properties such as being Ptolemaic \cite{HMS24}.
Furthermore, they naturally generalize so called tree-based single-rooted phylogenetic networks \cite{FS15} which have attracted a considerable amount of attention in the literature \cite{FF20,FSS18,H16,SGH24,Z16}. However in contrast to certain single-rooted phylogenetic networks 
not much is known about them in terms of encodings  whereby we mean that the network can be broken up into simpler combinatorial structures that, in a well-defined sense, uniquely determine the network (see e.g. \cite{GH12,HM13,vIM14} for such encodings). Arguably, the main attraction of them is that they  not only naturally lend themselves as a first port of call for developing reconstruction algorithms for arboreal networks so that their power can be exploited by evolutionary biologists, but also that they are generally easier to infer from real biological data.

As suggested  by the example in Figure~\ref{fig:example}, any answer to the question of what can be said about encodings for arboreal networks $N$  must in one way or another incorporate the roots of $N$ and also its {\em reticulation vertices} that is, the vertices of indegree two or more and outdegree one. Motivated by the fact that a rooted phylogenetic tree is sometimes constructed from an unrooted phylogenetic tree by including an outgroup for rooting purposes, we address the former by associating to an arboreal network $N$ with leaf set $X$ its planted version $N^p$ by adding for each of the roots $r$ of $N$ a new vertex $\hat{r}$ and an arc $(\hat{r},r)$.  Denoting the set of newly added vertices   by $R(N)$ then it is immediately clear  that the underlying graph $U(N^p)$ of $N^p$ is an unrooted phylogenetic tree with leaf set $X\cup R(N)$. Writing $U(N)$ for $U(N^p)$ to keep notation at bay, we address the latter by recording the reticulation vertices of $N$ in terms of a map  $\nu_N$ from the vertex set of $U(N)$ into a set $\{\bullet, \circ\}$ where a vertex is assigned $\bullet$ under $\nu_N$ if it is induced by a reticulation vertex of $N$ and $\circ$ otherwise.  
For the convenience of the reader, we depict for the arboreal network $N$ in Figure~\ref{fig:example}(i) the planting $N^p$ of $N$ and  the augmented tree $\mathcal U(N)=(U(N),\nu_N)$ associated to  $N$ in Figures~\ref{fig:example}(ii) and (iii), respectively.
Note that, from a conceptual level, such trees are unrooted phylogenetic trees $T$ along with a symbolic map from the vertex set of $T$ into a set of symbols which, in our case, are $\bullet$ and $\circ$ (see e.g. \cite{BD98,HH-RHMSW13,HMS19,HvIJJMMS24} and \cite[Section 7.6]{SS03} for other cases where such maps have been used in phylogenetics). Throughout the paper, we will refer to them as {\em augmented trees}. 


As is well-known, unrooted phylogenetic trees are encoded by collections of unrooted phylogenetic trees on 4 leaves called quartet trees that satisfy certain properties (see e.g. \cite{CS81,DE03,GHMS08,SS03}). 
Referring to a set of augmented quartet trees as an {\em enhanced quartet tree system}, we present in Theorem~\ref{the:main1} an encoding of augmented trees in terms of enhanced quartet tree systems that satisfy six non-trivial mutually independent properties. Since each one of them involves at most five elements of $X$, where $X$ is the union of the leaf sets of the quartet trees, it is immediately clear that it can be decided in order $O(|X|^5)$ time if an enhanced quartet tree system encodes an augmented tree or not. For arboreal networks the situation is less clear in that we present a non-trivial forbidden configuration that prevents even a binary augmented tree
from giving rise to an arboreal network. We conclude with showing that the situation improves considerably in case, in addition to the augmented tree, the leaves that are the outgroups and therefore correspond to the roots in the sought after arboreal network are known
(Theorem~\ref{the:main2}).

The paper is organized as follows. In the next section, we present formal definitions of the concepts that allow us to establish Theorem~\ref{the:main1}. In Section~\ref{sec:conditions}, we introduce our aforementioned six properties for enhanced quartet tree systems and show that they are independent of each other. Subsequent to that section, we then establish some basic results about enhanced quartet tree systems which will allow us to prove Theorem~\ref{the:main1} in Section~\ref{sec:theo1}. In Section~\ref{sec:arboreal}, we then present our forbidden configuration and prove Theorem~\ref{the:main2}.
  
\section{Preliminaries} 
\label{sec:prelims}
 From now on, $|X|\geq 4$. For a map $f:A\to B$ from a  set $A$ into a set $B$ of integers, we define the {\em support} $supp(f)$ of $f$ to be the set $\{a\in A\,:\, f(a)\geq 0\} $. 
 
 Suppose for the following that $G$ is a simple graph. Then we denote the vertex set of $G$ by $V(G)$ and the set of edges of $G$ by $E(G)$. Furthermore, we denote an edge between two distinct vertices $u$ and $v$ of $G$ by $\{u,v\}$. We call a vertex of $G$ of degree one a {\em leaf} of $G$ and denote the set of leaves of $G$ by $L(G)$. 
 
 Suppose that $v\in V(G)$. Then we call
 $v$ an {\em interior vertex} of $G$ if $v$ is not a leaf of $G $. If $u$ is a further interior vertex of $G$  such that $e=\{u,v\}\in E(G)$ then we call $e$ an  {\em interior edge} of $G$.
 If the degree of $v$ is two then we call $v$ a {\em subdivision point} of  $G$.  If $|V(G)|\geq 3$ and $x$, $y$, and $z$ are three vertices of $G$ such that $v$
 is the unique vertex that simultaneously lies on the path from $x$ to $y$, on the path from $x$ to $z$ and on the path from $z$ to $y$, then we call $v$ the {\em median of $x$, $y$, and $z$}. In this case, we also write $med_G(x,y,z)$ for $v$ where the order of $x$, $y$ and $z$ is of no relevance.  Note that a graph might contain three vertices that do not have a median -- see e.\,g.\,\cite{BvV99,BSH22,CHKKM08,HHMSS23,KM99,M11,M78} for more on medians in graphs. 
 
 \subsection{Phylogenetic trees}
  A {\em phylogenetic tree (on $X$)} is an unrooted  tree $T$ with no subdivision points and leaf set $X$.  Suppose for the following that $T$ is a phylogenetic tree on $X$. If $T$ is such that every interior vertex of $T$ has degree three, then we call $T$ a {\em binary} phylogenetic tree and if $T$  has a single interior vertex then we call $T$ a {\em star tree}. 
  We refer to a set $L$ of leaves of $T$ with at least two elements as a {\em multi-cherry} of $T$  if there exists a unique vertex in $T$ that is adjacent with every leaf in $L$. For a phylogenetic tree $T$ on $X$ and a non-empty subset $Y\subseteq X$ with at least four elements, we denote by $T|_Y$ the phylogenetic  tree obtained by first taking the span  $T(Y)$ of the elements in $Y$ and then suppressing all vertices of degree two in $T(Y)$. We say that two phylogenetic trees $T$ and $T'$ on $X$ are {\em equivalent}  if there exists a bijection 
 $\psi:V(T)\to V(T')$ that gives rise to a graph isomorphism between $T$ and
 $T'$ that is the identity on $X$.
 If $T$ is binary and has four leaves then we call $T$ a {\em quartet tree}. We denote the set of all quartet trees whose leaf set is contained in $X$ by $\cQ(X)$.
 We call a non-empty subset 
 $\cQ'\subseteq \cQ(X)$  of quartets trees a {\em quartet tree system (on $\bigcup_{q\in  \cQ'} L(q)$).}
 
 Suppose for the following that $q$ is a quartet tree in $\cQ(X)$ with leaf set $\{a,b,c,d\}$. 
  Let $s$ and $t$ denote the two interior vertices of $q$. If  the path from $a$ to $b$ crosses $s$ but not $t$ and the path from $c$ to $d$ crosses $t$ but not $s$ then we also denote $q$ by $ab|cd$ or, alternatively, by $cd|ab$
where, in each case, the order of $a$ and $b$ and $c$ and $d$ does not matter. Suppose that $q$ is a quartet tree in $\cQ(X)$ and that $T$ is a phylogenetic tree on $X$. Then we say that $q$ is {\em displayed} by $T$ if $q$ is equivalent with $T|_{L(q)}$. We denote the set of all quartet trees displayed by $T$ by $\cQ(T)$. 

\begin{figure}[h]
			\begin{center}
			\includegraphics[scale=0.45]{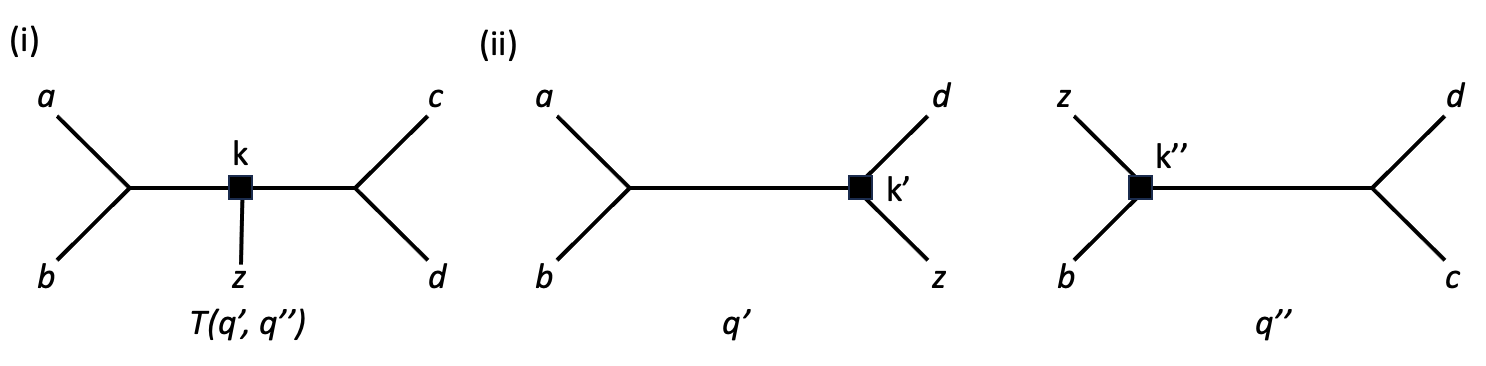}
				\end{center}
		\caption{
			\label{fig:support}
			(i) The phylogenetic tree  $T(q',q'')$ on $X=\{a,b,c,d,z\}$ that simultaneously displays the two quartet trees  $q' = ab|dz$ and $q'' = bz|dc$ in (ii). The vertices $k$, $k'$ and $k''$ are the median of the vertices $b$, $z$, and $d$ in the respective trees and are indicated as squares.	}
	\end{figure}

For two quartet trees  $q' = ab|dz$ and $q'' = zb|dc$, we denote the unique binary phylogenetic tree on $\{a,b,c,d,z\}$ that simultaneously displays $q'$ and $q''$ by $T(q',q'')$ -- see
Figure~\ref{fig:support} for an illustration. Note that the five quartet trees that make up the system $\cQ(T(q',q''))$ of quartet trees displayed by $T(q',q'')$ can be obtained by  taking, for example, the so-called semi-dyadic closure $scl_2(q',q'') $ of $q'$ and $q''$  -- see \cite{CS81} and \cite{SS03} for more on the semi-dyadic closure of a quartet tree system.

\subsection{Phylogenetic trees and splits}

A {\em partial split (on $X$)} is a set $\sigma=\{A,B\}$ where $A$ and $B$ are two non-empty disjoint subsets of $X$. If $A\cup B=X$, then we call $\sigma$ a {\em split} of $X$.
 In either case, we sometimes also write $A|B$ (or, equivalently, $B|A$) for $\sigma$ and  call $A$ and $B$ the {\em parts} of $\sigma$. 
 %
  If one of the parts of a partial split $\sigma$ on $X$ has size one then we call $\sigma$ a {\em trivial partial split} of $X$. For a subset $Y\subseteq X$ and a split $\sigma=A|B$ on $X$ such that $B'=Y\cap B\not=\emptyset$ and $ A'=Y\cap A\not=\emptyset$, we call the split $A'|B'$ on $Y$ the {\em restriction} of $\sigma$ to $Y$. We refer to a set $\Sigma$  of partial  splits on $X$ as a {\em system of partial splits (on $X$)}, and as a {\em split system  (on $X$)}
 if all elements of $\Sigma$ are splits on $X$. We say that a split $\sigma$ on $X$ {\em displays} a partial split $\sigma'$ on $X$ if there exists a subset $Y\subseteq X$ of $X$ such that $\sigma'$ is the restriction of $\sigma$ to $Y$.

Clearly, deleting from a phylogenetic tree $T$ on $X$ an edge $e$ induces a split $\sigma_e$ on $X$ given by taking the leaf set of one of the resulting connected components to be one part of $\sigma_e$ and the leaf set of the other to be the other part of $\sigma_e$. We denote the split system obtained from $T$ by taking the union of the splits $\sigma_e$, $e\in E(T)$ by $\Sigma(T)$. We say that a split $\sigma$ is {\em displayed} by a phylogenetic tree $T$ if $\sigma\in\Sigma(T)$. In case $T$ is a quartet tree, then we call the split induced by deleting the interior edge of $T$ the {\em quartet split} of $T$. We say that a split $\sigma$ of $X$ {\em displays} a quartet tree $q\in\cQ(X)$ if the quartet split associated to $q$ is displayed by $\sigma$.  Finally, for a non-trivial split $\sigma$ of $X$, we denote by $\cQ(\sigma)$ the set of quartet trees in $\cQ(X)$ that are displayed by $\sigma$.

Intriguingly, a phylogenetic tree can be reconstructed from its induced set of quartet trees in polynomial time. To make this more precise, we follow \cite{SS03} and denote for a quartet tree system $\cQ$ on $X$ the split system obtained by taking all splits $A|B$ of $X$ such that, for all pairwise distinct $a_1,a_2\in A$ and $b_1,b_2\in B$ we have  $a_1a_2|b_1b_2\in \cQ$ by $\Sigma(\cQ)$.
 By \cite[Theorem 6.3.7]{SS03}, it follows for any phylogenetic tree $T$ on $X$ that the split system $\Sigma(\cQ(T))$ associated to the set $\cQ(T)$ of quartet trees displayed by $T$ is the set of non-trivial splits displayed by $T$. Furthermore, $\Sigma(\cQ(T))$ and therefore also $T$ can be constructed from $\cQ(T)$ in polynomial time.

\subsection{Augmented trees and enhanced quartet tree systems}
\label{sec:augmented-trees}



Suppose for the following that $T$ is a phylogenetic tree on $X$. Then we call a map
$\nu:V(T)\to\{\circ, \bullet\}$ an {\em augmentation map (for $T$)} if, 
for all $v\in V(T)$,  we have that $\nu(v)\in\{\circ, \bullet\}$ 
if $v$ is an interior vertex of  $T$ and, otherwise, $\nu(v)=\circ$ holds. We call a vertex $v$ of $T$ with 
$\nu(v)=\bullet$ an {\em augmentation vertex} of $T$ under $\nu$ and denote the set of augmentation vertices of $T$ by $\mathcal A_{(T,\nu)}$, or $\mathcal A_T$, for short.  If $\nu$ is an augmentation map for $T$, then  we call a pair $\mathcal T=(T,\nu)$ an
 {\em augmented (phylogenetic) tree (on $X$)} and refer to $T$ as the {\em underlying tree of $\mathcal T$}. Note that a phylogenetic tree is an augmented tree without any augmentation vertices. 
 
 Extending the notion of an augmentation map for a phylogenetic tree to a quartet tree system $\cQ$, we call a map $\nu: V(\cQ) :=\bigcup_{q\in\cQ} V(q)\to \{\bullet,\circ\}$ an {\em augmentation map for $\cQ$} if, for all $v\in V(\cQ)$, we have that  $\nu(v)=\circ$ if $v$ is a leaf of a quartet tree in $\cQ$ and, otherwise,  $\nu(v)\in \{\bullet, \circ\}$.

Note that, in general, an augmentation vertex in an augmented tree $(T,\nu)$ can become a degree two  vertex on the path joining the two interior vertices of a quartet tree in $\cQ(T)$. To reflect this, we next associate to an augmentation map 
for the quartet trees in $\cQ(X)$ a map $\gamma:\cQ(X)\to \mathbb Z_{q\geq -1}$ 
which we call a  {\em subdivision map}.  We call a
tree $q'$ obtained from a quartet tree $q\in\cQ(X)$ by adding $k>0$ subdivision points to the interior edge  of $q$ a {\em subdivided} quartet tree. Furthermore, we call a subdivision point in a subdivided quartet tree an {\em augmentation point} of the quartet tree. 

We conclude this section with formalizing a central concept already mentioned in the introduction. We call a pair $(\gamma,\nu)$ an {\em enhanced quartet tree system (on $X$)} if  
$\gamma:\cQ(X)\to \mathbb Z_{\geq -1}$ is a subdivision map for the quartet trees in $\cQ(X)$ such that $supp(\gamma)\not=\emptyset$ and $\nu:V(supp(\gamma))\to\{\bullet,\circ\}$ is an augmentation map for $supp(\gamma)$.
As is straight-forward to check, an  augmented tree $\cT$ on $X$ whose underlying tree $T$ is not a star tree gives rise to an enhanced quartet tree system $(\gamma_T, \nu_T)$ on $X$ as follows.
 \begin{enumerate}
 	\item[$\bullet$] 
 	$\gamma_T : \cQ(X) \to \Z_{\geq -1} $ 
 	is the map 
 	which maps each quartet tree $q\in \cQ(X)$ to $-1$ if $q\not\in\cQ(T)$  and if 
 	$q\in \cQ(T)$ and $x$, $y$, $t$, $z$ are distinct elements in $X$ such that $q=xy|tz$ then
 	$\gamma_T(q)$ is the number of  augmentation vertices of $T$ that are interior vertices on the path in $T$ connecting 
 	the path from $x$ to $y$ and the path from $t$ to $z$. 
 	\item[$\bullet$] $\nu_T:V(\cQ(T))\to \{\circ,\bullet\}$ is the map defined by putting $\nu_T(v)=\bullet$ if there exists a quartet tree $q$ in $\cQ(T)$ such that $v$ is induced by a vertex in $\mathcal A_{\cT}$ and $\nu_T(v)=\circ$ otherwise. 
 	\end{enumerate}

\section{Six useful properties and their independence} 
 \label{sec:conditions}
 
 In this section, we introduce six useful properties called (A1)-(A6)
 for enhanced quartet tree systems. As we shall see in Section~\ref{sec:theo1}, they will allow us to characterize augmented trees in terms of such systems. To be able to state Condition~(A4), we require a further concept which we introduce next.
 
Suppose that $(\gamma,\nu)$ is an enhanced quartet tree system  on $X$ and that
 $q=ab|cd$ is a quartet tree in $supp(\gamma)$. Let $s$ be an augmentation point of $q$. Then we say that $s$ is \textit{supported} by $supp(\gamma)$ if
 there exist quartet trees $q'=bz|cd$ and $q'' = ab|dz$ in  $supp(\gamma)$ such that $q$ is displayed by $T=T(q',q'')$ so that $s$ gives rise to the median $k=med_T(b,z,d)$  in $T$, the median  $k'=med_{q'}(b,d,z)$ in $q'$ is induced by $k$ when restricting $T$ to  $\{b,c,d,z\}$, 
 the median $k''=med_{q''}(b,d,z)$ in $q''$ is induced by $k$ when restricting $T$ to  $\{a,b,d,z\}$, and 
 $\nu(k') =\nu(k'') =\bullet$. 
 
 To illustrate this concept consider the example depicted in Figure~\ref{fig:support} and 
 the enhanced quartet tree system $(\gamma,\nu)$ given by the subdivision map
 $\gamma:\cQ(X) \to \mathbb Z_{\geq-1}$ with $supp(\gamma)
 =\cQ(T(q',q''))$ and augmentation map $\nu:V(supp(\gamma))\to \{\bullet,\circ\}$ given, for all $v\in V(supp(\gamma))$ by $\nu(v)=\bullet$ if $v$ is induced by $k$  and $\nu(v)=\circ$ otherwise. Then $k$ is supported by $supp(\gamma)$ because it is an augmentation point of the quartet tree $ab|cd$ and $ab|cd\in \cQ(T(q',q''))$.

Suppose again  that  $(\gamma,\nu)$ is an enhanced quartet tree system  on $X$. Then, Properties (A1)-(A6) are as follows.
\begin{enumerate}
\item[(A1)] 
For all $a,b,c,d\in X$, at most one of the values $\gamma(ab|cd)$, $\gamma(ac|bd)$, and $\gamma(ad|bc)$ is non-negative.
\item[(A2)]  For all $x\in X-\{a,b,c,d\}$, if
 $\gamma(ab|cd)\geq0$ then
$$\mbox{$\gamma(ab|cx)\geq0$ and $\gamma(ab|dx)\geq0$}$$
or
$$\mbox{$\gamma(ax|cd)\geq0$ and $(bx|cd)\geq0$.}$$
\item[(A3)] If $q$ and $q'$ are quartet trees in $supp(\gamma)$ that share three pairwise distinct leaves $x$, $y$, and $z$, then 
$\nu(med_q(x,y,z))=\nu(med_{q'}(x,y,z))$. 
%
%
%
\item[(A4)] Every augmentation point of  a  quartet tree in $supp(\gamma)$ is supported by  $supp(\gamma)$.
\item[(A5)] Suppose $a,b,c,d,e \in X$ are such that
 $\gamma(ab|cd)>\gamma(ab|ce)\geq 0$ holds. 
If $\nu(med_{ab|ce}(a,c,e))=\bullet$ then
\begin{equation}\label{eqn: a5i}
	\gamma(ae|cd) = \gamma(ab|cd) - \gamma(ab|ce) - 1
\end{equation}
and, otherwise,
\begin{equation}\label{eqn: a5ii}
	\gamma(ae|cd) = \gamma(ab|cd) - \gamma(ab|ce)
\end{equation}
\item[(A6)] Suppose $a,b,c,d,e \in X$ are such that both $\gamma(ab|cd)\geq 0$ and $\gamma(bc|de) \geq 0$ hold. If $\nu(med_{ab|cd}(b,c,d)) = \bullet$ then
\begin{equation}\label{eqn: a6i}
	\gamma(ab|de) = \gamma(ab|cd) + \gamma(bc|de) + 1
\end{equation}
and, otherwise, 
\begin{equation}\label{eqn: a6ii}
	\gamma(ab|de)=\gamma(ab|cd) + \gamma(bc|de)
\end{equation}
%
%
%
\end{enumerate}

Before we start our investigation of Properties (A1)-(A6) with establishing independence relationships,  we remark that if $(\gamma,\nu)$ is an enhanced quartet tree system such that $\nu$ assigns the value $\circ$ to every vertex in $V(supp(\gamma))$, then $\gamma$ is the all-zero map and (A3) and (A6) always hold. Furthermore, Properties~(A4) and (A5) hold vacuously and  Properties~(A1) and (A2) are similar to the antisymmetry, symmetry and substitution properties used in \cite[Propositon 2]{BD86} to characterize quaternary relations on $X$ that are tree-like in the sense that there exists a potentially unresolved $X$-tree that induces that relation on $X$. In the context of this it should be noted that an $X$-tree is a generalization of a phylogenetic tree in that it is a tree $T$ along with a labelling $\phi:X\to V(T)$ such that every vertex $v$ of $T$ of degree one or two must be contained in $\phi(X)$. On the other hand and although $\gamma$ is similar in spirit to an object called a quartet weight function introduced in \cite{GHMS08}, it cannot be used to characterize edge-weighted phylogenetic trees as $\gamma$  does not satisfy \cite[Theorem 1]{GHMS08}.

\subsection{The independence of each of the Properties~(A1) - (A3) from the other five}

To see that Property~(A1) is independent of (A2) - (A6), consider the set $X = \{a,b,c,d\}$ and the map $\gamma:\cQ(X) \to \mathbb Z_{\geq -1}$ given by putting $\gamma(q) = 0$, for all $q\in \cQ(X)$. Furthermore, let $\nu:V(supp(\gamma))\to \{\bullet, \circ\}$ be the map given by $\nu(v)=\circ$, for all $v\in V(supp(\gamma))$. Then it is straight-forward to see that (A1) does not hold whereas all of (A2) - (A6) hold vacuously.

To see that Property~(A2) is independent of (A1) and  (A3) - (A6), consider the set $X = \{a,b,c,d,e\}$ and the map $\gamma:\cQ(X) \to \mathbb Z_{\geq -1}$ given by putting $\gamma(ab|cd) = 0$ and $\gamma(q)=-1$, for all other $q\in\cQ(X)$. Furthermore, let $\nu:V(supp(\gamma))\to \{\bullet, \circ\}$ be the map given by $\nu(v)=\circ$ if $v\in V(\{ab|cd\})$. 
Then (A1) holds because $|supp(\gamma)| = 1$, and
(A3) - (A6) hold vacuously.

To see that Property~(A3) is independent of (A1), (A2), and  (A4) - (A6), consider the set $X = \{a,b,c,d,e\}$ and the map $\gamma:\cQ(X) \to \mathbb Z_{\geq -1}$ given by putting $\gamma(ab|cd) = \gamma(ab|ce) = \gamma(ab|de) = 0$ and $\gamma(q)=-1$, for all other $q\in\cQ(X)$.
Furthermore, let $\nu:V(supp(\gamma))\to \{\bullet, \circ\}$ be the map given by 
$\nu(med_{ab|ce}(a,b,c)) = \bullet$ and $\nu(v)=\circ$, for all other $v\in V(supp(\gamma))$.
Then it is again straight-forward to  check (A1) and (A2) hold. Furthermore, (A4) - (A6) hold vacuously.

\subsection{Property~(A4) is independent of  Properties~(A1) - (A3), (A5), and (A6)}
We start the discussion of the independence of Property~(A4) from (A1) - (A3), (A5), and (A6), by remarking that if (A2), (A3) and (A6) hold then, using the notation in the definition of supporting an augmentation point, we can only have that (A4) does not hold if at least one of the quartet trees $q'$ and $q''$ is not contained in $supp(\gamma)$ or, if they are both contained in $supp(\gamma)$, that $med_{q'}(k')$ (and therefore also $med_{q''}(k'')$) is assigned $\circ$ under $\nu$.

To see the independence in the first case, 
consider the set $X=\{a,b,c,d\}$, the map $\gamma:\cQ(X)\to \mathbb Z_{\geq -1}$ given by $\gamma(ab|cd)=1$  and $\gamma(q)=-1$ for all other $q\in \cQ(X)$, and the map $\nu:V(\{ab|cd\})\to \{\bullet,\circ\}$ given by $\nu(v) =\circ$ for all 
$v\in V(\{ab|cd\})$. Then  it is straight-forward to see that Properties~(A1) - (A3) and (A5) and (A6) all hold. However,  (A4) does not hold because there exist no two quartet trees $q',q''\in supp(\gamma)$ such that $ab|cd$ is displayed by $T(q',q'')$.

In the second case, consider the set $X = \{a,b,c,d,e\}$ 
and the map $\gamma:\cQ(X) \to \mathbb Z_{\geq -1}$ given by putting 
$\gamma(ab|cd) =\gamma(ae|cd)=\gamma(be|cd)=1$, $\gamma(ab|ce)=\gamma(ab|de)=0$, and $\gamma(q) = -1$, for all other quartet trees $q\in\cQ(X)$. Furthermore, let $\nu : V(supp(\gamma)) \rightarrow \{\bullet,\circ\}$ be the map given by putting  $\nu(v) = \circ$, for all $v\in V(supp(\gamma))$. Then it is straight-forward to show that Properties~(A1) - (A3)  and (A5) and (A6) all hold. However, (A4) does not hold  in view of the definition of the map $\nu$.

\subsection{Property~(A5) is independent of Properties~(A1) - (A4) and (A6)}

To see that (A5) is independent of (A1) - (A4) and (A6), we distinguish for two quartet trees
$ab|cd$ and $ab|ce$ in $supp(\gamma)$ for which $\gamma(ab|cd)>\gamma(ab|ce)\geq0$ 
holds between the cases that $med_{ab|ce}(a,c,e)$
is assigned $\circ$ or $\bullet$ under $\nu$.  We start with the case
$\nu(med_{ab|ce}(a,c,e))=\bullet$.

\begin{figure}[h]
	\centering
	\includegraphics[scale=0.35]{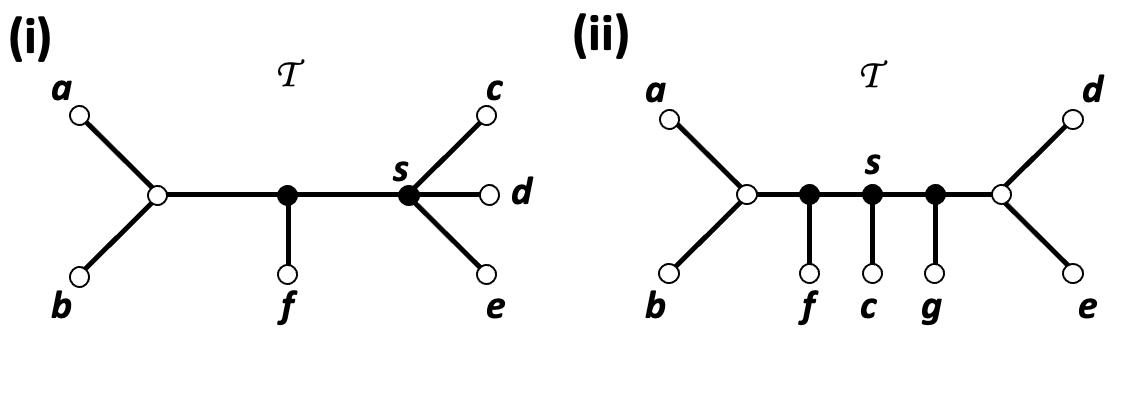}
	\caption{
		(i) The augmented tree $\cT$ considered in the discussion of the case that Property~(A5) is independent of Properties~(A1) - (A4) and (A6) for the case of Equation~(\ref{eqn: a5i}).  
		(ii) 	The augmented tree $\cT$ considered in the discussion of the case that Property~(A6) is independent of Properties~(A1) - (A5) for the case of Equation~(\ref{eqn: a6i}).  
		\label{fig:a5i-does-not-hold}
	}
\end{figure}

Consider the set $X = \{a,\ldots,f\}$ and the map $\gamma:\cQ(X)\to \mathbb Z_{\geq -1}$ whose support is given by the set of quartet trees displayed by the underlying tree $T$ of  the augmented tree $\cT$ depicted in Figure~\ref{fig:a5i-does-not-hold}(i). 
Put differently, $supp(\gamma)$ consists of the 
quartet trees $ab|xy$ with $x,y\in\{c,d,e,f\}$ distinct and, for all $x,y\in \{e,c,d\}$ distinct, of the quartet trees of the form $ab|xy$, or $bf|xy$ or $af|xy$.
Put $\gamma(ab|cd) = 1$, $\gamma(q)= 0$ for all other quartet trees in $\cQ(T)$, and $\gamma(q) = -1$ for all other $q\in\cQ(X)$.  Furthermore, let $\nu : V(supp(\gamma)) \rightarrow \{\bullet,\circ\}$ be the indicated  augmentation map. Put differently,
$\nu(v)=\circ$ if $v$ is a leaf of a quartet tree in $supp(\gamma)$ or $v= med_{ab|xy}(a,b,x)$ for some $x,y\in\{c,d,e,f\}$ distinct, and $\nu(v)=\bullet$ for all other $v\in V(supp(\gamma))$. 

The definition of $\nu$ combined with the fact that $\cQ(T)=supp(\gamma)$ implies immediately that Properties~(A1) - (A4) hold. To see that Property~(A6) holds, it suffices to check the 
quartet tree pairs $q$ and $q'$ where $q$ is of the form $ab|fx$  with $x\in \{c,d\}$ and $q'$ is of the form $q'=af|yz$ with $y,z\in\{c,d,e\}$ distinct. Since the vertex adjacent with $f$ in a quartet tree in $supp(\gamma)$ is assigned the value $\bullet$ under $\nu$   and $\gamma(ab|cd)=1$, the definition of $\gamma$ on the remaining quartet trees in $\cQ(T)$ implies that Equation~(\ref{eqn: a6i}) in (A6) holds.
To see that Equation~(\ref{eqn: a5i}) in Property~(A5) does not hold, note that $\gamma(ab|cd) =1>0= \gamma(ab|ce)$. Since $\nu(med_{ab|ce}(a,c,e))=\bullet$  Equation~(\ref{eqn: a5i}) implies that $-1=\gamma(ae|cd) = \gamma(ab|cd) - \gamma(ab|ce) - 1=0$ which is impossible.

To see that Property~(A5) is independent of (A1) - (A4) and (A6) if $\nu(med_{ab|ce}(a,c,e))=\circ$ holds, consider the set $X = \{a,b,c,d,e,f\}$ and the
augmented tree $\cT$ depicted in Figure~\ref{fig:a5i-does-not-hold}(i) where we replace the augmentation of the vertex $s$  in the augmentation map of $\cT$ by $\circ$.  Let $\gamma:\cQ(X)\to \mathbb Z_{\geq -1}$ be defined as before, and let $\nu:V(supp(\gamma))\to \{\bullet, \circ\}$ be the thus modified augmentation map. By definition, Properties~(A1) - (A4) hold. Using the same arguments as before, it follows 
that Property~(A6) also holds. To see that Equation~(\ref{eqn: a5ii}) in Property~(A5) does not hold, note again that $\gamma(ab|cd) > \gamma(ab|ce) \geq 0$. Since  $\nu(med_{ab|ce}(a,c,e))=\circ$  Equation~(\ref{eqn: a5ii}) implies that $-1=\gamma(ae|cd) = \gamma(ab|cd) - \gamma(ab|ce) =1$ which is impossible.

\subsection{Property~(A6) is independent of Properties~(A1) - (A5)}

To see that Property~(A6) is independent of (A1) - (A5), we distinguish for two quartet trees
$ab|cd$ and $bc|de$ in $supp(\gamma)$ between the cases that $med_{ab|cd}(b,c,d)$
is assigned $\circ$ or $\bullet$ under $\gamma$.  We start with the case
$\nu(med_{ab|cd}(b,c,d))=\bullet$.

Consider the set $X = \{a,\ldots,g\}$ and the augmented tree $\cT$ on $X$ depicted in
Figure~\ref{fig:a5i-does-not-hold}(ii).
 Let $T$ be the underlying tree of $\cT$ and let $\gamma:\cQ(X)\to \mathbb Z_{\geq -1}$ denote the map with $supp(\gamma)=\cQ(T)$ given by putting $\gamma(q)=1$ if $q\in \cQ'=\{ab|de, ab|cd, ac|de, bc|de, ab|ce\}$, $\gamma(q)=0$ if $q\in \cQ(T)-\cQ'$, and
$\gamma(q)=-1$ if $q\in \cQ(X)-\cQ(T)$. Furthermore, let $\nu:V(supp(\gamma))\to \{\bullet, \circ\}$ be the indicated  augmentation map. Put differently, $\nu(v)=\circ$ if $v$ is a leaf of a quartet tree in $supp(\gamma)$, or there exists a quartet tree of the from $ab|xy$ with $x,y\in\{c,d,e,f,g\}$ distinct  and $v=med_{ab|xy}(a,b,x)$, or there exists a quartet tree of the from $xy|de$ with $x,y\in\{a,b,c,f,g\}$ distinct  and $v=med_{xy|de}(d,e,x)$.
Furthermore, $\nu(v)=\bullet$ for all other $v\in V(supp(\gamma))$. 

Clearly, Properties~(A1) - (A3) hold. Also, it is straight-forward to check that 
 Properties~(A4) and (A5) hold. Furthermore, $\nu(med_{ab|cd}(b,c,d))=\bullet$.
 However, Equation~(\ref{eqn: a6i}) in (A6) does 
 not hold because $\gamma(ab|de)=1\not=3=\gamma(ab|cd)+\gamma(bc|de)+1$.



To see that (A6) is independent of (A1) - (A5) if $\nu(med_{ab|cd}(b,c,d))=\circ$, consider again the set $X =\{a,\ldots,g\}$ and the augmented tree $\cT$ on $X$ depicted in 
Figure~\ref{fig:a5i-does-not-hold}(ii)
but with the augmentation of the vertex $s$ replaced by $\circ$ in that augmentation map of $\cT$. Let $\cQ'$ and $\gamma:\cQ(X)\to \mathbb Z_{\geq -1}$ be defined as before. Finally, let $\nu: V(supp(\gamma))\to \{\bullet,\circ\}$ be the thus modified  augmentation map. Then, as before, (A1) - (A5) hold. Furthermore, $\nu(med_{ab|cd}(b,c,d))=\circ$. However, Equation~(\ref{eqn: a6ii}) in (A6) does not hold because $\gamma(ab|de)=1\not=2=\gamma(ab|cd)+\gamma(bc|de)$.



\section{Basic properties of enhanced quartet tree systems}

To be able to relate enhanced quartet tree systems  with splits which we do next, we 
start with the following result
whose Part~(ii) is adapted from \cite[Lemma 1]{GHMS08}.

\begin{lemma}\label{lem:A7}
	Suppose that $(\gamma,\nu)$ is an enhanced quartet tree system on $X$ that satisfies Properties~(A1) and (A2).  
	\item[(i)] If $a,b,c,d,e\in X$ are such that $\gamma(ab|ce), \gamma(ab|de) \geq 0$, then $\gamma(ab|cd) \geq 0$.
	\item[(ii)] If $(\gamma,\nu)$ also satisfies Properties~(A3), (A5), and (A6), then  the following property also holds:
\begin{enumerate}
\item[(A7)] For all $a,b,c,d,e \in X$, $\gamma(ab|cd)\geq \min\{\gamma(ab|ce),\gamma(ab|de)\}$.
\end{enumerate}
\end{lemma}
\pf 
(i) Assume for contradiction that $\gamma(ab|cd) = -1$. 
Applying (A2) to $ab|ce$ and $d$ implies that both $\gamma(ab|cd) \geq 0$ and $\gamma(ab|de) \geq 0$ or that $\gamma(ad|ce)\geq 0$ and $\gamma(bd|ce)\geq 0$. Since, by assumption, $\gamma(ab|cd) = -1$, it follows that $\gamma(ad|ce) \geq 0$. Similarly, applying (A2) to $\gamma(ab|de)$ and $c$ implies that $\gamma(ac|de) \geq 0$. But this is impossible in view of (A1).

(ii) Suppose that Properties~(A3), (A5) and (A6) also hold but that (A7) does not hold, that is,
\begin{equation}\label{eqn:A7}
	\gamma(ab|cd) < \min\{\gamma(ab|ce),\gamma(ab|de)\}
\end{equation}
We claim first that
$$\gamma(ab|cd) \geq 0$$
To see the claim, assume that $\gamma(ab|cd) = -1$. Then Inequality~(\ref{eqn:A7}) implies that $\gamma(ab|ce) \geq 0$ and $\gamma(ab|de) \geq 0$. Applying (A2) to $\gamma(ab|ce) \geq 0$ and $d$  and noting that $\gamma(ab|cd) = -1$, we obtain $\gamma(bd|ce) \geq 0$. Similarly, applying (A2) to $\gamma(ab|de) \geq 0$ and $c$, yields $\gamma(bc|de) \geq 0$, a contradiction to Property~(A1).
Hence, $\gamma(ab|cd) \geq 0$ as claimed. By Inequality~(\ref{eqn:A7}), $\gamma(ab|de)> 0$.

Using Inequality~(\ref{eqn:A7}), we obtain $\gamma(ab|ce) > \gamma(ab|cd) \geq 0$. Note that, by Property~(A3), $\nu(med_{ab|cd}(a,c,d)= \nu(med_{ad|ce}(a,c,d))$.  Hence, by (A5), we have $\gamma(ad|ce) = \gamma(ab|ce) - \gamma(ab|cd)$ if $\nu(med_{ab|cd}(a,c,d)) = \circ$ and $\gamma(ad|ce) = \gamma(ab|ce) - \gamma(ab|cd) - 1$, otherwise.  Thus,
$\gamma(ad|ce)\geq0$. Since $\gamma(ab|cd) \geq 0$ and therefore $\gamma(ad|bc) = -1$ because of Property~(A1), we obtain 
$
\gamma(bd|ce) \geq 0$ 
by applying (A2) to $\gamma(ad|ce)$ and $b$. 
In view of  (A3), $\nu(med_{ab|cd}(b,c,d))= \nu(med_{bd|ce}(b,c,d))$ and so, by (A6),
 $$
 \gamma(ab|ce)=\gamma(ab|cd)+\gamma(bd|ce)
 $$
  holds if $\nu(med_{ab|cd}(b,c,d)) = \circ$. Combined with Property~(A3),   it follows that
 $\nu(med_{bd|ce}(b,d,e))=\nu(med_{ab|de}(b,d,e))=\nu(med_{ab|cd}(b,c,d))$. By Property~(A6),
 $$
 \gamma(ab|ce)=\gamma(ab|de)+\gamma(bd|ce)
 $$ 
 must hold too in this case. 
  Similar arguments imply that 
 $\gamma(ab|de)+\gamma(bd|ce)+1=\gamma(ab|ce)=\gamma(ab|cd)+\gamma(bd|ce)+1$
  holds in case $\nu(med_{ab|cd}(b,c,d)) = \bullet$.
In either case, $\gamma(ab|de) = \gamma(ab|cd)$ follows, contradicting Inequality~(\ref{eqn:A7}).
\epf

\begin{lemma}\label{lem:split}
	Suppose $(\gamma,\nu)$ is an enhanced quartet tree system that satisfies Properties~(A1)-(A3), (A5), and (A6). If $ss'|tt'$ is a quartet tree in $supp(\gamma)$ 
	and $A,B$ are disjoint subsets of $X$ such that 
	$s, s' \in A$, $t,t' \in B$, $Q(A|B)\subseteq supp(\gamma)$, and $|A|+|B|$ is maximum, then 
	$A|B$ is a split of $X$.
	\end{lemma}
\pf
Suppose that the elements $s$, $s'$, $t$ and  $t'$ and the sets $A$ and $B$ are  as in the statement of the lemma and assume for contradiction that $A|B$ is not a split of $X$.  Then there is an element $x\in X-(A\cup B)$. 
We claim that there exist 
(not necessarily distinct) elements  $a_1,a_2,a_3\in A$ and 
$b_1,b_2,b_3\in B$ with $|\{a_1,a_2,b_1,b_2\}|=4$ such that 
\begin{align*}
	\label{observation1}
	\mbox{$\gamma(a_1a_2|b_3x) = -1$ and $\gamma(a_3x|b_1b_2) = -1$.}
\end{align*}
Indeed, if no such elements $a_1,a_2,a_3\in A$ and 
$b_1,b_2,b_3\in B$ existed then, since $A\cap B=\emptyset$ we have that either $A\cup\{x\}$ and $B$ are disjoint subsets of $X$ or  $A$ and $B\cup\{x\}$ are disjoint subsets of $X$. Without loss of
generality, we may assume that $A'=A\cup\{x\}$ and $B$ are disjoint subsets of $X$. Since, by assumption, $\gamma(aa'|bb') \geq 0$ for all $a,a'\in A$ and $b,b'\in B$, it follows that $\gamma(aa'|bb') \geq 0$ for all $a,a'\in A'$ and $b,b'\in B$. 
Since $ss'|tt'\in Q(A'|B)$ and $|A|+ |B| <|A'|+|B|$ this is impossible as  $|A|+|B|$  is maximum. Thus, the claim holds.

Since, by (A7), $\gamma(a_1a_2|b_3x) \geq \min \{\gamma(a_1a_2|b_3b),\gamma(a_1a_2|bx)\}$ for all $b \in B - \{b_3\}$ and $\gamma(a_1a_2|b_3x)=-1$, it follows for all such $b$ that $\gamma(a_1a_2|bx) = -1$ because $a_1a_2|b_3b\in Q(A|B)\subseteq supp(\gamma)$.
Similarly, $\gamma(ax|b_1b_2) = -1$ for all $a \in A-\{a_3\}$. With $a = a_1$ and $b = b_1$, we obtain $\gamma(a_1a_2|b_1x) =- 1$ and $\gamma(a_1x|b_1b_2) =-1$. But this is impossible since $\gamma(a_1a_2|b_1b_2) \geq 0$ and (A2) holds for $\gamma(a_1a_2|b_1b_2)$ and $x$. Hence, $A|B$ is a split of $X$.
\epf

To be able to  prove our next result,  we require further notation. 
Suppose that $(\gamma,\nu)$ is an enhanced quartet tree system and that $\sigma$ is a split contained in $\Sigma(supp(\gamma))$. 
Then we denote by  $\cQ(\sigma)^-\subseteq \cQ(\sigma)$ the set of all quartet trees  $q\in \cQ(\sigma)$ such 
that $\sigma$ is the sole split in $\Sigma(supp(\gamma))$ that displays $q$. 
To help illustrate this definition, we return to the enhanced quartet tree system $(\gamma,\nu)$ that we
used  to illustrate the concept of an augmentation point being supported by $supp(\gamma)$. Then since $\sigma=\{a,b\}|\{c,d,z\}$ is the sole split that displays the quartet tree $ab|dz$, we have $ab|dz \in \cQ(\sigma)^-$. However $ab|cd\not\in\cQ(\sigma)^-$ because $ab|cd$ is also displayed by the split $\{a,b,z\}|\{c,d\}$.

\begin{lemma}\label{lem:gamma-tight-display}
	Suppose $(\gamma,\nu)$ is an enhanced quartet tree system  on $X$ that satisfies Properties~(A1) - (A6).
	If 	 $\sigma$ is a split of $X$ such that $Q(\sigma)\subseteq supp(\gamma)$, then $\gamma(q)=0$, for all $q\in \cQ(\sigma)^-$. 
	\end{lemma}
\pf
Suppose that $\sigma=A|B$ is a split of $X$ such that $Q(\sigma)\subseteq supp(\gamma)$.
Assume for contradiction that there exists some quartet tree $q=ab|cd\in \cQ(\sigma)^-$ such that $\gamma(q)\geq 1$. Then $q$ has at least one augmentation point $s$. Without loss of generality, we may assume that $s$ is such that there is no further augmentation point of $q$ between $s$ and $med_q(b,c,d)$. By Property~(A4), $s$ is supported by $supp(\gamma)$. Hence, 
there exists some $z\in X-\{a,b,c,d\}$ and  quartet trees $q' = ab|dz$ and $q''=bz|cd$ in $supp(\gamma)$ such that $q$ is displayed by $T=T(q',q'')$, the median $med_T(b,z,d)$  in $T$ induces the median  $s'=med_{q'}(b,d,z)$ in $q'$ when restricting $T$ to  $\{a,b,d,z\}$
and the median $s''=med_{q''}(b,d,z)$ in $q''$ when restricting $T$ to  $\{b,c,d,z\}$, and $\nu(s')=\nu(s'')=\bullet$.

Let $A'$ and $B'$ be disjoint subsets of $X$ such that $a,b\in A'$ and $d,z\in B'$, $\cQ(A'|B')\subseteq supp(\gamma)$ and $|A'|+|B'|$ is maximum. Furthermore, let $A''$
and $B''$ be disjoint subsets of $X$ such that $b,z\in A''$ and $c,d\in B''$, $\cQ(A''|B'')\subseteq supp(\gamma)$ and $|A''|+|B''|$ is maximum.
Note that such sets $A'$ and $B'$ must exist because $q'\in supp(\gamma)$
and that $A''$ and $B''$ must exist because $q''\in supp(\gamma)$. Since $(\gamma,\nu)$ satisfies Properties~(A1) - (A6), it follows by Lemma~\ref{lem:split}
that $A'|B'$ and $A''|B''$  are splits of $X$. Since, by construction, $q\in\cQ(A'|B')$ 
and $q\in \cQ(A''|B'')$ and either $z\in A$ or $z\in B$ must hold because $\sigma$ is a split of $X$,  at least one of $A'|B'$ 
and $A''|B''$ cannot be $\sigma$, a contradiction because  $q\in \cQ(\sigma)^-$.
\epf

\section{Characterizing augmented trees}
\label{sec:theo1}
One of the fundamental results in phylogenetics is the 
 Split-Equivalence Theorem \cite[Theorem 3.1.4]{SS03} -- see also \cite{B71} which may be stated as follows. Saying that a split system $\Sigma$ is {\em compatible} if for any two distinct 
splits $\sigma$ and $\sigma'$ in $\Sigma$  there exists a part $A\in \sigma$ and a part $A'\in\sigma'$ such that $A\cap A'=\emptyset$ then that theorem guarantees that
for any split system $\Sigma$ on  $X$ that
contains all trivial splits on $X$, there is a phylogenetic tree $T$ on $X$ with $\Sigma=\Sigma(T)$ if and only 
if $\Sigma$ is compatible. Moreover, if such a phylogenetic tree  $T$ exists, then, up to equivalence, $T$ is unique.  Our first main result (Theorem~\ref{the:main1}) may be viewed as its analogue. To be able  to state it, we require further concepts. We say that two augmented trees $(T,\nu)$ and $(T',\nu')$ are {\em augmentation-equivalent} if $T$ and $T'$ are equivalent as phylogenetic trees and, with $\psi:V(T)\to V(T')$ denoting the underlying bijection, we have for all $v\in V(T)$ that $\nu(v)=\nu'(\psi(v))$.

\begin{theorem} \label{the:main1}
	Suppose that $|X|\geq 5$ and that $(\gamma,\nu)$ is an enhanced quartet tree system on $X$. Then there exists an augmented tree $\cT=(T,\nu)$ such that $T$ is not the star tree, $\gamma=\gamma_T$ and $\nu=\nu_T$ if and only if 
	$(\gamma,\nu)$ is an enhanced quartet tree system on $X$ that satisfies Properties~(A1)-(A6). 
	Moreover, if such an augmented tree $\cT$ exist then, up to augmentation-equivalence, $\cT$ is unique.
	\label{unrootedthm}
\end{theorem}

\pf First suppose that $\cT$ is an augmented tree on $X$. Let $T$ denote the underlying tree of $\cT$. Then,  the pair $(\gamma_T,\nu_T)$ introduced in 
Section~\ref{sec:augmented-trees}  is an enhanced quartet tree system on $X$.
 We need to show that  $(\gamma_T,\nu_T)$ satisfies Properties~(A1) - (A6).

To see that  $(\gamma_T,\nu_T)$ satisfies Property~(A1), suppose that $a$, $b$, $c$, and $d$ are pairwise distinct elements in $ X$. Since  a phylogenetic tree on $X$ can display at most one quartet tree on $Y=\{a,b,c,d\}$,  it follows that at most one quartet tree on $Y$ can be assigned a non-negative value under $\gamma_T$. Thus, $(\gamma_T,\nu_T)$ satisfies (A1).

To see that $(\gamma_T,\nu_T)$ satisfies Property~(A2), suppose that
$a,b,c,d\in X$ are such that  $ab|cd $ is a quartet tree in $supp(\gamma_T)$. Then $\Sigma(T)$ 
contains a split $\sigma=A|B$ 
that displays $ab|cd$. Without loss of generality, we
may assume that $a,b\in A$ and $c,d\in B$. Let $x\in X-\{a,b,c,d\}$. 
Then either $x\in A$ or $x\in B$.
If $x\in A$, then $\sigma$ displays the quartet trees $ax|cd$ and $bx|cd$. Hence,
$ax|cd\in supp(\gamma_T)$ and $bx|cd\in supp(\gamma_T)$. Since similar arguments apply if $x\in B$,
it follows that  $(\gamma_T,\nu_T)$ satisfies (A2). 

To see that $(\gamma_T,\nu_T)$ satisfies Property~(A3), suppose that 
$q$ and $q'$ are quartet trees in $supp(\gamma_T)$ that share three pairwise distinct leaves $x$, $y$, and $z$. Then the medians $med_q(x,y,z)$ 
and $med_{q'}(x,y,z)$ in $q$ and $q'$, respectively, 
are induced by the median $med_T(x,y,z)$ in $T$. Hence,
$\nu_T(med_q(x,y,z))=\nu_T(med_{q'}(x,y,z))$. Thus,
$(\gamma_T,\nu_T)$ satisfies (A3).

To see that $(\gamma_T,\nu_T)$ satisfies Property~(A4), suppose that $s$ is an augmentation  point of a quartet
tree $q=cd|az$ in $supp(\gamma_T)$. Then $s$ must be an augmentation vertex of $T$. Since an augmentation point of $q$ can, in $T$, only be adjacent to at most one leaf of $T$ and $|X|\geq 5$, there must exist interior vertices $v$ and $v'$ of $T$ such that $\{s,v\}$ and $\{s,v'\}$ are interior edges of $T$. 
Choose a leaf $b$ of $T$ such that the path from $s$ to $b$ neither crosses $v$ nor $v'$. Then  $q'=cd|bz$ and $q''=bc|az$ are quartet trees that are displayed by $T$. Hence, $q', q''\in supp(\gamma_T)$. Since, by the Splits-Equivalence Theorem mentioned above, $\Sigma(T)$ is compatible, it follows that every split in $\Sigma(T)$ that displays $q'$ also displays $q$. Furthermore and by the same argument, we also have that a split in $\Sigma(T)$ displays $q$ if and only if it displays $q'$ but not $q''$. Hence, $T(q',q'')$ is induced by restricting $T$ to $\{a,b,c,d,z\}$. Thus, since $q$ is displayed by $T$ we also have that
$q$ is also displayed by $T(q',q'')$. In addition,
the median $med_T(b,c,z)$  in $T$ induces the median  $s'=med_{q'}(b,c,z)$ in $q'$ when restricting $T$ to  $\{b,c,d,z\}$ and the median
$s''=med_{q''}(b,c,z)$ in $q''$ when restricting $T$ to  $\{a,b,c,z\}$. Since $s$ is an augmentation vertex of $T$, it follows that
$\nu_T(s') =\nu(s'') =\bullet$.
Hence, $s$ is supported by $supp(\gamma_T)$. Thus, $(\gamma_T,\nu_T)$ satisfies (A4).

To show that $(\gamma_T,\nu_T)$ satisfies Property~(A5), suppose $a,b,c,d,e\in X$ are such that the quartet trees
$ab|cd$ and $q=ab|ce$ are both contained in $supp(\gamma_T)$ and that 
$\gamma_T(ab|cd)>\gamma_T(q)\geq 0$.  
Then similar arguments as in the proof that $(\gamma_T,\nu_T)$ satisfies Property~(A4) imply that $T(q,q')$ is induced by restricting $T$ to 
$\{a,b,c,d,e\}$. It follows that the median $v=med_q(a,c,e)$ in $q$ is induced by the median $med_{T(q,q')}(a,c,e)$ in $T(q,q')$ which is itself induced by the median $w=med_T(a,c,e)$ in $T$. Similarly, the median  $v'=med_{q'}(a,c,e)$ in $q'$ is induced by $w$.
Thus,  if $w$ is not an augmentation vertex of $T$ then
$\nu_T(v)=\nu_T(v')=\circ$ and if $w$ is an augmentation vertex of $T$ then $\nu_T(v)=\nu_T(v')=\bullet$. Since $w$ is an interior vertex on the path in $T$ that joins the path
from $a$ to $b$ in $T$ and the path from $c$ to $d$ in $T$, Equation~\ref{eqn: a5i} follows if 
$\nu_T(v)=\bullet$ and if $\nu_T(v)=\circ$ then  Equation~\ref{eqn: a5ii} follows.
Thus, $(\gamma_T,\nu_T)$ satisfies (A5).

%

Finally, to see that $(\gamma_T,\nu_T)$ satisfies Property~(A6), suppose that there 
exist elements $a,b,c,d,e\in X$ such that the quartet trees 
$q=ab|cd$ and $q'=bc|de$ are both contained in $supp(\gamma_T)$. 
Then there exists splits $\sigma=A|B$ and $\sigma'\in\Sigma(T)$ such that $\sigma$ displays $q$ and $\sigma'$ displays $q'$. Without loss of generality, we may assume that $a,b\in A$ and $c,d\in B$. Note that $e\not\in A$ because otherwise $q'\in supp(\gamma_T)$ would not hold. Hence, $e\in B$. In particular, the number of augmentation vertices in the interior of the path in $T$ separating the path from $a$ to $b$ from the path from $e$ to $d$ equals $\gamma_T(q)+\gamma_T(q')+1$ if $w=med_T(b,c,d)\in \mathcal A_T$ as in this case
$\nu_T(med_{q}(b,c,d))=\bullet =\nu_T(med_{q'}(b,c,d))$. If $w
\not\in\mathcal A_T$ then $\gamma_T(q)+\gamma_T(q')$
follows because in this case  $\nu_T(med_{q}(b,c,d))=\circ =\nu_T(med_{q'}(b,c,d))$.
Hence
$(\gamma_T,\nu_T)$ satisfies (A6).

In summary, it follows that  $(\gamma_T,\nu_T)$ is an enhanced quartet tree system  on $X$ that satisfies Properties~(A1) - (A6).

To prove the converse, suppose that $(\gamma,\nu)$ is an enhanced quartet tree system  on 
$X$ that satisfies Properties~(A1) - (A6). We prove the theorem by induction on $|supp(\gamma)|$. The base case is $|supp(\gamma)|= 1$. Let $q\in supp(\gamma)$. Then $q$ has $i$ augmentation vertices, some $i\in\{0,1,2\}$ but, in view of (A4),
no augmentation points. Since, for all $i\in\{0,1,2\}$, we have that $q$ along with its augmentation vertices is an augmented tree such that $\gamma=\gamma_q$, and $\nu=\nu_q$, the theorem holds.

Assume for the remainder that $|supp(\gamma)|\geq 2$ and that the induction hypothesis holds for  all enhanced quartet tree systems $(\gamma',\nu')$ on $X$ with $|supp(\gamma')|<|supp(\gamma)|$.  
Let $p=aa'|bb'$ be a quartet tree in $supp(\gamma)$ such that $\gamma(p)$ is minimum. In view of (A4), it follows that $\gamma(p)=0$. Let $A$ and $B$ be two disjoint subsets of $X$ such that $a,a'\in A$, $b,b'\in B$, $\cQ(A|B)\subseteq supp(\gamma)$ and $|A|+|B|$ is maximum.  Note that such subsets must exist since $p\in supp(\gamma)$. By Lemma~\ref{lem:split}, $\sigma=A|B$ is a split of $X$. Choose a quartet tree $q^*=ss'|tt'$ in $\cQ(\sigma)^-$. Since $\cQ(\sigma)\subseteq supp(\gamma)$ it follows by Lemma~\ref{lem:gamma-tight-display} that $\gamma(q)=0$ holds for all $q\in \cQ(\sigma)^-$.

 We next associate  an 
 enhanced quartet tree system $(\gamma',\nu')$ on $X$ to $(\gamma,\nu)$ and show that $(\gamma',\nu')$ satisfies the  induction hypothesis. 
 To this end, we  first define a map  $\mu:V(supp(\gamma))\to \{\bullet, \circ\}$ as follows.
 Let $v\in V(supp(\gamma))$. Then we put $\mu(v) =\bullet$ if there exists a quartet tree $q\in \cQ(\sigma)^-$ such that $v$ is an interior vertex of $q$ and at least one of the two interior vertices of $q$ is assigned the value $\bullet$ under $\nu$. Otherwise, we put $\mu(v)=\nu(v)$.
 To complete our definition of $(\gamma',\nu')$, 
we put, for all $q\in \cQ(X)$, 
$\gamma'(q)=-1$ if  $q\in \cQ(\sigma)^-$ or $\gamma(q)=-1$. Otherwise, we define $\gamma'(q)$ to be  the number of augmentation points of $q$ under $\mu$.
Clearly, $(\gamma',\mu)$ is an enhanced quartet tree system 
because $|supp(\gamma)|\geq 2$. Furthermore, since every augmentation point of a quartet tree in $supp(\gamma')$ is also an augmentation point of a quartet tree in $supp(\gamma)$
we have $supp(\gamma')\subseteq supp(\gamma)$. In view of this, we define 
$\nu':V(supp(\gamma')) \to \{\bullet,\circ\}$ as the restriction $\mu|_{V(supp(\gamma'))}$
of $\mu$ to $V(supp(\gamma'))$.

Note that $(\gamma',\nu')$ is also an enhanced quartet tree system on $X$ because 
$|supp(\gamma)|\geq 2$, and that $supp(\gamma')\subseteq supp(\gamma)$ also holds
for $(\gamma',\nu')$. To help keep notation at bay, we will from now on only mean $\nu'$ when referring to an augmentation point/vertex of a quartet tree in $supp(\gamma')$. 
Note also that  $q^*\not \in supp(\gamma')$
because  $q^*\in \cQ(\sigma)^-$. Hence, $|supp(\gamma')|<|supp(\gamma)|$.
We next show that $(\gamma', \nu')$ satisfies Properties~(A1)-(A6).

Clearly, $(\gamma', \nu')$ satisfies Property~(A1) because $(\gamma, \nu)$ satisfies (A1).
To see  that $(\gamma',\nu')$ satisfies Property~(A2), assume for contradiction that 
there exist elements 
$a,b,c,d\in X$ with $ab|cd\in supp(\gamma')$ but (A2) does  not hold. Then, for 
some $x\in X-\{a,b,c,d\}$, $i\in \{c,d\}$, and $j\in \{a,b\}$, we have 
\begin{eqnarray}
ab|ix, jx|cd\not\in supp(\gamma').
\label{ijeqn}
\end{eqnarray}
Without loss of generality, we may assume that $i=c$ and that $j=b$.
Then $ab|cx, bx|cd\not\in supp(\gamma')$.
Since $ab|cd\in supp(\gamma')\subseteq supp(\gamma)$ and $(\gamma,\nu)$ satisfies (A2), 
$$
\mbox{$ab|cx\in supp(\gamma)$ and  $ab|dx \in supp(\gamma)$}
$$
or 
$$
\mbox{$ax|cd\in supp(\gamma)$ and $bx|cd\in supp(\gamma)$}
$$
must hold.
The definitions of $\gamma'$ and $\nu'$ combined with Assumption~(\ref{ijeqn}) imply 
that  $ab|cx$ and  $bx|cd$ are both contained in $Q(\sigma)^-$. But $ab|cx\in Q(\sigma)^-$  implies that $x$ and $c$ are contained in the same part of $\sigma$ and  $bx|cd
\in Q(\sigma)^-$  implies that $x$ and $c$ are not contained in the same part of $\sigma$; a contradiction. Thus, $(\gamma',\nu')$ satisfies (A2).

To see that $(\gamma',\nu')$ satisfies Property~(A3), let $q,q'\in supp(\gamma')$
be such that there exist three pairwise distinct leaves $x,y,z\in L(q)\cap L(q')$. Put  $v=med_q(x,y,z)$ and $w=med_{q'}(x,y,z)$. Then $\nu(v)=\nu(w)$ because $(\gamma,\nu)$ satisfies (A3). Furthermore, $\nu'(v)\not=\nu'(w)$ can only hold if and only if for one of $v$ and $w$, say $v$, we have $\nu'(v)=\bullet$ and $\nu'(w)=\nu(w)=\nu(v)=\circ$.
By the definition of $\nu'$, it follows that there exists a quartet tree $q''\in \cQ(\sigma)^-$
 such that $v$ is an interior vertex of $q''$ and the other interior vertex of $q''$ is assigned the value $\bullet$ under $\nu$. By the definition of  $v$, it follows that $x,y,z\in L(q'')$. But then $w$ is also an interior vertex of $q''$ and, so,
 $\nu'(w)=\bullet$, a contradiction.   Thus, $(\gamma',\nu')$ satisfies (A3).


To see that $(\gamma',\nu')$ satisfies Property~(A4), suppose there exists some  quartet tree $q=ab|cd\in supp(\gamma')$ that has an augmentation point $s$. Since $supp(\gamma')\subseteq  supp(\gamma)$, Property~(A4) applied to $(\gamma,\nu)$ implies that $s$ is also an augmentation point of $q$ and that $s$  is supported in $supp(\gamma)$.
Hence,  there exist quartet trees $q'=bz|cd$ and $q'' = ab|dz$ in  $supp(\gamma)$ such that $q$ is displayed by $T=T(q',q'')$, the median $med_T(b,z,d)$  in $T$ induces the median  $k'=med_{q'}(b,d,z)$ in $q'$ when restricting $T$ to  $\{b,c,d,z\}$,
$k''=med_{q''}(b,d,z)$ is the median in $q''$ when restricting $T$ to  $\{a,b,d,z\}$, and 
$\nu(k') =\nu(k'') =\bullet$. 

If neither $q'$ nor $q''$ is a quartet tree in $\cQ(\sigma)^-$ then  $\nu'(k')=\nu(k')=\bullet=\nu(k'')=\nu'(k'')$ and so (A4) holds for $(\gamma',\nu')$ 
in this case.  So assume that one of $q'$ and $q''$ is a quartet tree in $\cQ(\sigma)^-$. Without loss of generality, assume that $q'\in \cQ(\sigma)^-$. Then  $\nu'(k')=\bullet$  by the definition of $\nu'$, and $q''\not\in \cQ(\sigma)$. Hence,  $\nu'(k'') =\nu(k) =\bullet$ must also hold which, in turn, implies $\nu'(k')=\bullet=\nu'(k'')$. Thus, $(\gamma',\nu')$ satisfies (A4) also in this case.

To see that $(\gamma',\nu')$ satisfies Property~(A5), suppose for contradiction that there exist
 $a,b,c,d,e \in X$ such that  $ab|cd$ and $q=ab|ce$ are quartet trees that are both contained in $ supp(\gamma')$ and that  $0\leq \gamma'(q)<\gamma'(ab|cd)$.

 We first claim that 
$0\leq \gamma(q)<\gamma(ab|cd)$. 
Since  $0 \leq \gamma'(q)\leq \gamma(q)$ clearly holds, it suffices to show that $\gamma(q)<\gamma(ab|cd)$ holds.
Assume for contradiction that $\gamma(ab|cd)\leq \gamma(q)$.
Then since,  for any quartet tree $p\in supp(\gamma')$, the definition of $\gamma'$
implies that the difference  $\gamma(p)-\gamma'(p)$ is zero or one, it is straight forward 
to see that 
$\gamma'(q)=\gamma(q)-1$ and  $\gamma'(ab|cd)=\gamma(ab|cd)$ must hold. Hence, $q\in \cQ(\sigma)$ and, as a consequence, 
$ab|cd\not\in \cQ(\sigma)$.  Since $\sigma$ is a split of $X$, we also have 
$ad|ce\in \cQ(\sigma)$. Hence, $\gamma(ad|ce)\geq 0$. 
 By Property~(A6) applied to 
$ad|ce$ and $ab|cd$, it follows that $q$ is displayed by $T(ab|cd, ad|ce)$ and also that
$\gamma(ab|cd)<\gamma(q)$ in case  $\nu(med_{ab|cd}(a,c,d))=\bullet$ or if $\nu(med_{ab|cd}(a,c,d))=\circ$ and $\gamma(ad|ce)\geq 1$. But this case cannot hold as it implies   $\gamma(q)-1=\gamma'(q)<\gamma'(ab|cd)=\gamma(ab|cd)<\gamma(q)$ which is impossible.  Thus, $\nu(med_{ab|cd}(a,c,d))=\circ$ and $\gamma(ad|ce)=0$ must hold. But this is also impossible since the definition of $\nu'$ combined with the fact that 
$ab|cd$ and $q$ are contained in $\cQ(\sigma)$  implies that $\gamma(q)=\gamma'(q)=\gamma(q)-1$ which is impossible.
This completes the proof of the claim.

To complete the proof that $(\gamma',\nu')$ satisfies Property~(A5), assume first that we have 
$\nu'(med_q(a,c,e))=\circ$. Then $\nu(med_q(a,c,e))=\circ$ by the definition of $\nu'$. Since $(\gamma,\nu)$ satisfies Property~(A5), we obtain $\gamma(ae|cd)=
\gamma(ab|cd)-\gamma( q)$. Since   $ab|cd$ is displayed by $\sigma$ if and only if precisely one of $q$ and $ae|cd$ is displayed by $\sigma$ it follows that 
$\gamma'(ab|cd)=\gamma(ab|cd)-1$ if and only if $\gamma'(p)=\gamma(p)-1$ for a
single $p\in\{q,ae|cd\}$ and that  $\gamma'(p')=\gamma(p')$ for  $p'\in\{q,ae|cd\}-\{p\}$. Thus, 
$\gamma'(ae|cd)=\gamma'(ab|cd)-\gamma'( q)$ as required. 
 
 So assume that $\nu'(med_q(a,c,e))=\bullet$. Then, by the definition of $\nu'$, either   $\nu(med_q(a,c,e))=\bullet$ or  $\nu(med_q(a,c,e))=\circ$. If $\nu(med_q(a,c,e))=\bullet$ then 
similar arguments as in the previous case imply that $\gamma'(ae|cd)=
\gamma'(ab|cd)-\gamma'( q)-1$, again as required. 

So assume that $\nu(med_q(a,c,e))=\circ$. Then since $(\gamma,\nu)$ satisfies (A5), we obtain $\gamma(ae|cd)=\gamma(ab|cd)-\gamma( q)$. Note that $\gamma'(q)=\gamma(q)-1$. Consequently $\gamma(ae|cd)=\gamma'(ae|cd)$ because $q\in \cQ(\sigma)$ and so, by (A6) applied to $q$ and $ae|cd$, we obtain $ae|cd\not \in \cQ(\sigma)$. Thus   $\gamma'(ae|cd)=\gamma'(ab|cd)-\gamma'( q)-1$ again, as required. This completes the proof that $(\gamma',\nu')$ satisfies (A5).

To see that $(\gamma',\nu')$ satisfies Property~(A6), assume that $a,b,c,d,e\in X$ such that $q=ab|cd$ and $q'=bc|de$ are quartet trees in  $supp(\gamma')$. Then  $q$ and $q'$ are also contained in $supp(\gamma)$ 
because $0\leq \gamma'(p)\leq \gamma(p)$ holds for all $p\in\{q,q'\}$. Note that, independent of the value of $\nu(med_{ab|cd}(b,c,d))$, we have that $ab|de$ is displayed by $T(q,q')$.
For all splits $\sigma'$ of $X$, it follows that $\sigma'$ displays $ab|de$
if and only if $\sigma'$ displays one of $q$ and $q'$. 

Suppose first that  $\sigma$ does not display $ab|de$. Then $\gamma'(p)=\gamma(p)$, for all $p\in\{ab|de, q,q'\}$. Furthermore,
$\nu(med_{ab|cd}(b,c,d))= \nu'(med_{ab|cd}(b,c,d))$. Since $(\gamma,\nu)$ satisfies Property~(A6), if follows that $(\gamma',\nu')$ satisfies Property~(A6).

So assume that $\sigma$ displays $ab|de$. Without loss of generality, assume that 
$\sigma$ displays $q$. Then $\gamma'(q')=\gamma(q')$. 
If $\nu'(med_q(a,c,d))= \circ$
then $\nu(med_q(a,c,d))= \circ$ and $\gamma(p)=\gamma'(p)$, for all $p\in\{ab|de,q\}$. Hence, {\tiny }
$(\gamma',\nu')$ satisfies Property~(A6) because  $(\gamma,\nu)$ satisfies that property.

So assume that $\nu'(med_q(a,c,d))= \bullet$. Then, by the definition of $\nu'$, either $\nu(med_q(a,c,d))= \bullet$ or $\nu(med_q(a,c,d))= \circ$.
 In the former case,
$\gamma(ab|de)=\gamma'(ab|de)+1$ and $\gamma'(q)=\gamma(q)-1$.
Similar arguments as before imply that $(\gamma',\nu')$ satisfies Property~(A6) also in this case. Finally, if
$\nu(med_q(a,c,d))= \circ$, then $\gamma(ab|de)=\gamma'(ab|de)$ and $\gamma'(q)=\gamma(q)-1$. Since $(\gamma,\nu)$ satisfies Property~(A6), we obtain $\gamma'(ab|de)=\gamma(ab|de)=\gamma(q)+\gamma(q')=\gamma'(q) +\gamma'(q')+1$.
Hence, $(\gamma',\nu')$ satisfies Property~(A6) also in this case.


In summary, we have that $(\gamma',\nu')$ is an enhanced quartet tree system that satisfies Properties~(A1) - (A6).

Since $|supp(\gamma') | < |supp(\gamma) |$
and $(\gamma',\nu')$ satisfies (A1)-(A6) 
it follows by induction hypothesis that there exists an augmented tree $\cT'$ on $X$ with underlying tree $T'$ 
such that $\gamma_{T'}=\gamma'$ and 
$\nu_{T'}=\nu'$. By the Splits-Equivalence Theorem reviewed above, it follows that $\Sigma(T')$ is compatible. Note that, by construction, $\sigma\not\in\Sigma(T')$. But then $\Sigma(T')\cup\{\sigma\}$ must be compatible as otherwise there would exist four pairwise distinct elements  $x$, $y$, $z$, and $r$ in $X$ such that one of the four possible quartets trees with leaf set $\{x,y,z,r\}$ is contained in $supp(\gamma')-\cQ(\sigma)^-$ and the other is contained in $\cQ(\sigma)^-$. But this is impossible because both quartet tree systems are contained in $supp(\gamma)$  and $(\gamma,\nu)$ satisfies (A1). By the Splits-Equivalence Theorem,
it follows that there exists a phylogenetic tree $T$ on $X$ such that the quartet tree system $\cQ(T)$ induced by $T$ equals $supp(\gamma)$.

Let $e=\{u,w\}$ denote  the edge in $T$ whose deletion induces the split $\sigma$ on $X$.  Since $T'$ is clearly obtained from $T$ by collapsing $e$, we obtain $V(T)=V(T')-\{med_{T'}(s,s',t)\} \cup\{u,w\}$. Since
$med_T(s,s',t)$  and $med_T(s,t,t')$ are both incident with $e$, we may assume without loss of generality that $u=med_T(s,s',t)$ and that $w=med_T(s,t,t')$. Consider the augmentation map $\hat{\nu}:V(T)\to \{\bullet, \circ\}$ given, for all $v\in V(T)$, by putting 
$\hat{\nu}(v)=\nu_{T'}(v)$ if $v\in V(T')- \{med_{T'}(s,s',t)\}$ and 
$\hat{\nu}(v)= \nu(med_{q^*}(s,s',t))$ if $v=u$ and $\hat{\nu}(v)= \nu(med_{q^*}(t,t',s))$ 
if $v=w$. Since $\hat{\nu}$ induces $\nu_T$, $\nu'=\nu_{T'}$ and 
$\gamma'=\gamma_{T'}$, it follows by the definition of $\nu'$ that $\nu=\nu_T$ and 
$\gamma=\gamma_T$, as required.

The remainder of the theorem is a straight-forward consequence of the fact that any two phylogenetic trees $T$ and $T'$ such that $\cQ(T)$ equals $\cQ(T')$ must be equivalent, 
where by equal we mean that for every quartet tree $q$ in $\cQ(T)$ there must exist  a (necessarily unique) quartet tree in $\cQ(T')$ that is equivalent with $q$ and vice versa (see \cite[Corollary 6.3.8]{SS03}) and the  fact that for any augmented tree $\cT$ with underlying tree $T$ we must have that $\nu_T=\nu$ and $\gamma_T =\gamma$.
\epf

\section{Augmented trees and arboreal networks}
\label{sec:arboreal}

We now turn our attention to arboreal networks which we already described briefly in the introduction. Formally speaking, an {\em arboreal network $N$ (on $X$)}  is a multiple-rooted directed acyclic graph with leaf set $X$ and no vertices of indegree and outdegree one such that planting $N$ and ignoring directions results in a phylogenetic tree on $X\cup R(N)$. In this context, $R(N)$ denotes the set of vertices  that were attached via incoming arcs to the roots of $N$. Note that, by construction, $R(N)\not=\emptyset$. 
Clearly, every planted arboreal network $N$  be transformed back into the arboreal network  that gave rise to it by collapsing the newly added arcs.

Although every planted arboreal network gives rise to an augmented tree as described in the introduction, not every augmented tree gives rise to a planted arboreal network and, therefore, to an arboreal network. For example, an augmented tree can have interior edges $e$  such that $e$ has at least two augmentation points, whereas in an augmented tree induced by an arboreal network an edge can have at most one augmentation point. Furthermore since a reticulation vertex in an arboreal network has outdegree one, no augmentation vertex in an augmented tree can be adjacent with more than one leaf. 

\begin{figure}[h]
	\begin{center}
		\includegraphics[scale=0.40]{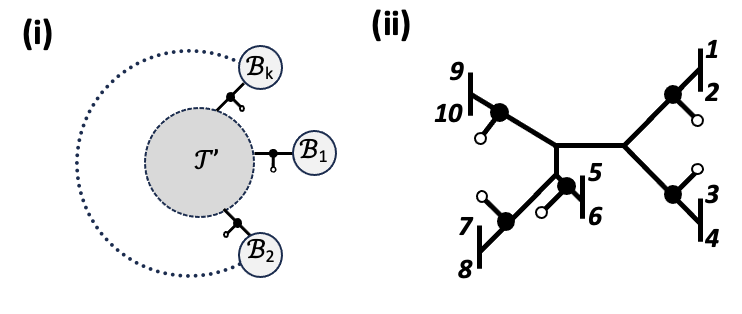}
	\end{center}
	\caption{\label{fig:corona}
(i) A configuration in a binary augmented tree $\cT$ on $X$ without adjacent augmentation vertices that prevents $\cT$ from being transformable into an arboreal network $N$ 
-- see text for details.  (ii) For $k=5$, an example of an augmented tree on $X=\{1,\ldots, 10\}$ in the form of the forbidden configuration in (i). For all $1\leq i\leq 5$, the disks $\mathcal B_i$ are cherries and $\mathcal T'$ is the augmented tree in the form of a path of length two all of whose vertices are not augmentation vertices. }
\end{figure}

Even though these  insights are undoubtedly useful, from the point of view of reconstructing an arboreal network $N$ from an augmented tree $\cT=(T,\nu)$  such that  $\cT$ is augmentation-equivalent with the augmented tree $\mathcal U(N)$ associated to $N$  they are however not sufficient,  even if $\cT$ is binary and does not contain augmented vertices that are adjacent with each other.  One of the reasons for this is the configuration depicted in Figure~\ref{fig:corona}(i) as a subgraph where the grey disk marked $\cT'$ is an augmented subtree of $\cT$ that contains no leaves of $\cT$
and the vertices indicated by a $\circ$ are leaves of $\cT$. For all  $3\leq i\leq k$, some integer  $k\geq 3$,
the disk marked $\mathcal B_i$ is the augmented tree induced by (i) attaching a new leaf $l$ to the vertex $v$ in the underlying tree $B_i$ that is adjacent with the augmentation vertex $w$ of $\cT$ that is not in $\mathcal B_i$, (ii) assigning the value $\circ$ to $l$ and preserving all other values under $\nu$, and (iii) deleting the edge 
$\{v,w\}$ from $T$.

More precisely,  assume that $\cT$ is a binary augmented trees such that no two augmentation vertices are adjacent and that contains the configuration in that figure as an augmented 
(induced) subtree of $\cT$. Then since $\cT$ is binary and a reticulation vertex in an arboreal network has outdegree one,
it is straight forward to see that $\cT'$ must contain a vertex that has outdegree three. Thus, there does not exist an arboreal network $N$ such that $\mathcal U(N)$ is augmentation-equivalent with $\cT$. On the other hand, if $\cT$ is such that one of the leaves in the configuration in Figure~\ref{fig:corona}(i) is an augmented  tree with at least two vertices all of which are non-augmentation vertices then it is straight-forward to see that there exists an arboreal network $N$ such that 
$\mathcal U(N)$ and $\cT$ are augmentation equivalent. However even in this case there does not
exist, in general, an arboreal network $N$ such that $\mathcal U(N)$ and $\cT$ are augmentation-equivalent if 
$\cT'$ contains at least one augmentation vertex.

We conclude this section with Theorem~\ref{the:main2}  which states that  there is an efficient algorithm (Algorithm~\ref{alg:check}) to check if there exists an arboreal network $N$ such that its associated augmented tree $\mathcal U(N)$ is augmentation-equivalent with $\cT$ provided the set of leaves of $\cT$ that corresponds to the set of roots of $N$ is know.  To establish it, we require further terminology. Let $\cT=(T,\nu)$ be an augmented tree on $X$ and let $Y\subseteq X$ be a non-empty subset of $X$. If there exists a single edge $e$ of $T$ such that when deleting $e$ the span $T(Y)$ of $Y$ is a connected component of $T$, then
we write $S(Y)$ for $T(Y)$ and call $\mathcal S(Y) = (S(Y), \nu|_{S(Y)})$ the  \textit{rooted augmented span} of $Y$. We denote the root of $S(Y)$ by $r_Y$. If $Y$ is such that $r_Y$ is an augmentation vertex  of $\cT$ and every vertex in $V(S(Y)) - \{r_Y\}$ is not an augmentation vertex of $\mathcal{T}$ then we call $\cS(Y)$ \textit{augmentation-rooted}.
If $r$ is an augmentation vertex of $\mathcal{T}$ and $Y$ is augmentation-rooted at $r$ then we also write $Y_r$ for $Y$.

\begin{algorithm}
	\caption{\label{alg:check}
	}
	\begin{algorithmic}[1]
		\Require {Binary augmented tree $\mathcal{T} = (T, \nu)$ on $X$ and a non-empty subset $R\subseteq X$.}
		\Ensure {Arboreal network $N$ on $X-R$ with root set $R$ such that $\mathcal U(N)$ and $\mathcal{T}$ are augmentation equivalent  or the statement ``No such network exists".}
		
		\State Let $\alpha$ be the number of augmentation vertices of $\mathcal{T}$.
		\If {$\alpha \geq |R|$ {\bf or} there exists an augmentation vertex that does not lie on a path between two elements in $R$ {\bf or} $\alpha\geq 2$ and there exists a root $r \in R$ such that there exist no further augmentation vertex $h$ and root $r' \in R$ such that $h$ is on the path joining $r$ and $r'$ } \Return ``No such network exists".
		
		\Else
		\State Put $Z = X \cup R$, $\alpha' = \alpha$, $T' = T$, $\nu' = \nu$, and $\mathcal{T'} = (T', \nu')$
		\ForAll {$r \in R$} direct their incident edge in $T'$ away from $r$.
		\EndFor
		\ForAll {$x \in X-R$} direct their incident edge in $T'$ towards $x$.
	 \EndFor
		
		\While {$\alpha' \neq 0$} 
		Choose a non-empty subsets $Y \subseteq Z$ such that $\mathcal{S}(Y)$ is augmentation rooted and direct all edges of $T'$ in $S(Y)$ so that the direction of the edges incident with the elements in $Y$ is respected and every interior vertex of $S(Y)$ has indegree 1 and outdegree 2. 
		\If {this is not possible} \Return ``No such network exists".
		\Else { direct the edge of $T'$ incident with $\rho_Y$ but not in $S(Y)$ so that $\rho_Y$ has indegree 2 and outdegree 1. Collapse $S(Y)$ into the vertex $\rho_Y$, replace
			$\nu'(\rho_Y)=\bullet$ by $\nu'(\rho_Y)=\circ$,  and preserve the value under $\nu'$ of all other vertices of the resulting phylogenetic tree. Put $\alpha'=\alpha'-1$ and $Z=Z- Y \cup \{\rho_Y\}$. Replace $\mathcal{T}'$ by the new augmented tree which we also call $\cT'$.}
		\EndIf
		\EndWhile
			\If {$T'$ can be directed such that the inherited directions are preserved and only vertices with indegree 1 and outdegree 2 are generated} reverse the collapsing of the trees $S(Y)$ in the while loop starting at Line 7, to obtain the tree $T$.
			\Else \State  \Return ``No such network exists".
			\EndIf
			\EndIf
			
		\end{algorithmic}
	\end{algorithm}
We remark in passing that since a binary augmented tree $\cT=(T,\nu)$ on $X$ has $2|X|-3$ edges and can have at most $|X|-2$ augmentation vertices, 
Algorithm~\ref{alg:check} is clearly efficient since, each time the while loop in Line~7 is entered,
 the number of augmentation vertices of $\cT$ is decremented by one and the number of edges of $T$ is reduced by at least one. 

\begin{theorem} \label{the:main2}
	Let $\cT$ be a binary augmented tree on  $X$ and let  $R\subseteq X $ be a non-empty subset of $X$. Then Algorithm~\ref{alg:check} 
	applied to $\cT$ and $R$ either returns an arboreal network $N$ on $X-R$ with root set $R$ such that $\mathcal U(N)$ is augmentation-equivalent with $\cT$ or the statement ``No such network exists".
\end{theorem}
\pf
A straight-forward inductive argument on the number $m_T$ of augmentation vertices of $\cT$ 
implies that the reversal of the collapsing sequence constructed in the while loop (Line~7) described in Line~10 generates an arboreal network $N$ whose associated augmented tree $\mathcal U(N)$ is augmentation-equivalent with $\cT$, should such a network exist. Central for this is the fact that, in Lines~5  and 6, directions are assigned to the edges incident with elements in $R$ and in $X-R$, respectively. The base case of the induction is $m_T=0$ which corresponds to the if-case in Line~10. Note that this line must be reached because $\cT$ can have at most $|X|-2$ augmentation vertices and the number of augmentation vertices is decremented by one each time the while loop is entered.
\epf

\section{Conclusion}
In this paper, we have studied the question of what can be said about encodings of arboreal networks, a certain kind of multiple rooted directed acyclic graph that might prove useful for extending phylogeny-based HGT-inference methods to the case where an overall species tree might be uncertain or unavailable or the information that the transfer happened between bacteria inhabiting different ecological niches is important and therefore should be preserved. Our results are combinatorial (Theorems~\ref{the:main1} and \ref{the:main2}) as well as algorithmical (Algorithm 1) in nature and also include a forbidden configuration that prevents a binary augmented tree $\cT$ from giving rise to an arboreal network.

Numerous questions however remain that might warrant further study. These include characterizing augmented trees that cannot be transformed, in the above sense, into an arboreal network in terms of forbidden configurations and, as a first step towards reconstructing augmented trees from real biological data and, in the long run, arboreal networks, estimating augmentation maps for quartet trees from data. 

\bibliographystyle{abbrv}
\bibliography{references}

\end{document}